# BLOGRANK: RANKING WEBLOGS BASED ON CONNECTIVITY AND SIMILARITY FEATURES

Apostolos Kritikopoulos
*Athens University of Economics and Business*
*Patision 76, Athens, Greece*
*+30 6977687978*
*apostolos@kritikopoulos.info*

Martha Sideri
*Athens University of Economics and Business*
*Patision 76, Athens, Greece*
*+30 2108203149*
*sideri@aueb.gr*

Iraklis Varlamis
*Athens University of Economics and Business*
*Patision 76, Athens, Greece*
*+30 2108203160*
*varlamis@aueb.gr*

## Abstract

*A large part of the hidden web resides in weblog servers. New content is produced in a daily basis and the work of traditional search engines turns to be insufficient due to the nature of weblogs. This work summarizes the structure of the blogosphere and highlights the special features of weblogs. In this paper we present a method for ranking weblogs based on the link graph and on several similarity characteristics between weblogs. First we create an enhanced graph of connected weblogs and add new types of edges and weights utilising many weblog features. Then, we assign a ranking to each weblog using our algorithm, BlogRank, which is a modified version of PageRank.*

*For the validation of our method we run experiments on a weblog dataset, which we process and adapt to our search engine. (http://spiderwave.aueb.gr/Blogwave). The results suggest that the use of the enhanced graph and the BlogRank algorithm is preferred by the users.*

## 1. Introduction

'Blogging' or 'web logging' is a popular way of publishing information on the web. Due to its ease of use, blogging has become a tool for web based communication, collaboration, knowledge sharing, reflection and debate. This results in a huge and important information base that expands dramatically fast [12].

According to the weblog search engine of Technorati (http://technorati.com/weblog) that tracks 27.2 million weblogs, the size of blogosphere doubles every 5.5 months [21] and today is 60 times bigger than 3 years ago. Although their structure is simple (text, images and links), and their content corresponds to personal beliefs, weblogs carry lots of information that can be useful to many scientific fields.

Analysis of the blogosphere [2] shows communities of weblogs which are type related, topic related or hyperlink connected. Every weblog consists of a series of entries, expressing uniform or contradictory opinions and link to other entries or web pages. Links are used as suggestions, as a means to express agreement or disagreement [16] or even in order to bias search engine results (i.e. spam weblogs, google bombs).

Apart from the explicit relation expressed by hyperlinks, weblogs can be related based on more implicit features, such as common topics: two weblogs with posts on common topics are considered relevant even though they may not be connected with a hyperlink. Another interesting feature of weblogs that is not examined thoroughly by researchers, is the participation and contribution of users. The same users participate in more than one weblogs, as owners, affiliated authors or commentators. This creates a kind of relation between the weblogs that is not necessarily supported by hyperlinks between entries.

The explicit hyperlinks between weblog entries, and the implicit links between weblogs that emanate from the similarity in the users that post or in the topics discussed, turn the blogosphere into an interesting graph. In this multi-coloured and weighted graph the nodes are either the weblogs or the contained posts and the edges are either hyperlinks or similarity links. The weights express the strength of relation between nodes.

Typical ranking methods based on simple link graphs (i.e. PageRank) favour A-list weblogs, with many reciprocal links [7]. Although the graph formed by the hyperlinks between weblog posts is part of the web graph, the ranking algorithms for web pages seem to be insufficient for the following reasons:

- The number of links between weblog entries is very small. Thus, the weblog entries graph is very sparse and the ranking algorithms do not perform well.

- Weblog-specific information is not exploited in its full extent.

The enhancement of the graph of the blogosphere with implicit links based on various weblog aspects, increases the density of the graph, and allows us to devise and experiment with a method for ranking the nodes of the enhanced graph in terms of importance. The proposed ranking method aims to fit a state of the art ranking algorithm to the enhanced weblog graph. For this reason:
- We process the weblog graph from a certain viewpoint that gives a denser graph.
- When a source node links to several target nodes, we decide on the possibility of a user to select one of the targets based on various criteria. The criteria represent similarity in topics and contributors between the source and target nodes, the freshness of the target node and the strength of connection between nodes.

In order to test the efficiency of our ranking method we attached a sample weblog dataset provided by Nielsen BuzzMetrics, Inc. to an experimental search engine and launched a weblog search service. However, the results of our work can be applied to any weblog collection and is on our next plans to do so with the Greek section of the blogosphere.

The next step was to evaluate user satisfaction for different ranking methods, using uninformed (blind) testing. The results presented to a user query, are ranked by one of the available ranking methods. The method was selected randomly each time. The evaluation was based on the posts selected and the order of selection. Since the users are not aware of the algorithm used for each query we are confident that the tests are totally unbiased. More details on the evaluation method are available at [11].

It is important to clarify that the work in this paper and mainly the experimental part has been adapted to the dataset we had available. The current implementation combines tools and methods, which we already have tested in other experiments and which we assembled and tuned in order to create a functional weblog search engine. However, we gain useful insight on the blogosphere its intrinsic features and it is on our next plans to extend this work outside of the dataset, which is a small portion of the whole weblog graph.

The rest of this paper is structured as follows: In section 2 we present existing work in weblog analysis. In section 3 we discuss the details of the dataset. We illustrate the structure of the graph formed by the posts and the explicit hyperlinks among them and give facts, from the statistical analysis, which prove the sparseness of the graph. In the same section we present our conception of the enhanced weblog graph, we give examples on the different types of implicit links and on the final result. Section 4 presents in detail the ranking algorithm, describes how the enhanced weblog graph has been exploited and gives the motive behind our decisions. Section 5 illustrates our experiments with the dataset and the search engine. In separate subsections we present how we extended the graph, give the top ranked weblogs and discuss the setup and results of the user evaluation process. Section 6 contains our conclusions and presents the future plans of our work in this area.

## 2. Related Work

The work presented in this paper is based on our previous experience in ranking nodes into highly connected graphs. The most significant output of our work in this direction is *SpiderWave* [10], a research search engine for the Greek portion of the Web (about 4 million documents, basically the .gr domain) designed by our research group. The engine can be reached from the Web site of our University (http://spiderwave.aueb.gr) and is an alternative search engine for Greek content. SpiderWave has an algorithm for ranking every page based only on the Greek fragment of the Web graph. For this paper we made available a test service for webblogs.

The service can be accessed at:

http://spiderwave.aueb.gr/Blogwave

The majority of ranking algorithms that detect authority nodes in an interconnected graph ([19], [15], [8]) reside on the density of the graph [14]. However, it is proved that the graph of weblog entries is not very connected and the use of a ranking algorithm such as PageRank is problematic. This is mainly due to the fact that weblog authors usually record their personal opinion on a topic without linking to the opinion of others. Even when links are provided, they usually point to newspaper articles and not to weblog entries. The same holds with the dataset we processed in our experiments. As can be seen in the statistics of section 3 (Table 1) we have only 0.27 edges per node in our graph.

Kurland & Lee [13] tackle the problem of non interconnected documents (i.e. non-hypertext documents) by inducing links between documents based on content similarity. Moreover, they produce directed links from the more to the less diverse documents. However, the analysis of content, which is a prerequisite in this approach, can prove very time consuming in the quickly evolving blogosphere.

Fujimura et al [5] notice the problem of the weakly interconnected weblog graph and add a reputation property into every node in the graph in order to bias the random surfer model. The EigenRumor algorithm is applied on the graph of weblog post and as output attaches hub, authority and reputation scores first to weblogs and authors and as a consequence to their posts. Trackback information is used in the experiments, which is not the case in our dataset. Additionally, our ranking method is applied on the weblogs' graph, which is a generalization of the posts' graph. The sparseness of the weblog graph has already been noticed by researchers and a number of implicit links have been created to increase the density of the graph. The implicit links suggested by Adar et al. [1] denote similarity between nodes in content and out-links. Although author name is not always characteristic on the author's identity (authors may have double identities inside a weblog, or the same username can be used in many weblogs by more than one user) if properly used can be a supportive factor on the similarity between weblogs. For weblogs in the same server the username is unique for each registered user [3]. The combination of weblog server name and user name can be consequently used as user id. As a result, the number of authors in common between two weblogs strengthens the implicit link between them.

The dataset we process contains links to non-weblog URLs (marked as Press). These are web pages' URLs and are ending nodes on our graph, since they have only incoming links. The ranking of such nodes is based on the number of incoming links and the authority of their referees. Such a ranking will show the most influential [6] non-weblog nodes for the weblogs community.

The result of running a ranking algorithm in the graph of weblog posts will be to find highly interconnected posts that have been referenced by many and refer to many posts. Although top ranked posts are very interesting for someone to read, they are nothing more than information bits, with few content. It would be interesting in an analysis of the blogosphere to find those weblogs that rank high in importance by collecting many influential posts. In order to achieve such a ranking we need to abstract the initial graph of posts to a graph of weblogs.

In the next section we summarize on the dataset we employed for our study and on the methodology we followed for ranking weblogs.

## 3. Structure

### 3.1 The dataset

In order to explain the rational of our approach, we briefly present the structure of the specific dataset we used and the features we employed in our analysis. Although the weblog structure is not standard, all weblogs share the following structure: A weblog contains one or more posts and has a URL. Every post consists of an author, a body, a date and time of publishing and a URL of the full, individual article (the permalink). A post optionally includes: comments of readers, category tags and links to referrers (trackback links). Trackback and comment information is unavailable from the test dataset.

Only the number of *comments* and *trackbacks* can be retrieved by processing the contents of each post. Since this type of information is not standard for all weblog servers the numbers can be retrieved for a small portion of the dataset (number of trackbacks for 0.26% and number of comments for 0.52% of the posts).

*Topic* information is available for 23.75% of the posts (for 1,8 million out of 7.6 millions unique weblog entries). The *author* name is not always very useful, mainly due to:

- anonymous posting. It is an option in several weblogs to allow users to post entries without providing a name.
- common user names. Several user names (i.e. admin, webmaster, john etc) appear in more than one weblogs but we can easily assume that correspond to different persons
- double identities. A person can have different usernames in different weblogs and less possible but probable to have two accounts in the same weblog.

Usually, members of a community contribute in more than one weblogs based on the interests they have in common. Even when there are no intrinsic links between two weblogs (hyperlinks, permalinks), an overlap between the contributor names indicates a relation. Bigger the overlap indicates stronger relation. As a result, we decide to employ author name information as a factor of relation between weblogs and we assign to the factor a weight of importance which can be changed upon case.

Similarly the choice of topic is subjective to the author. However, when combined with hyperlinks the analysis of topic and author information gives an important aspect of the blogosphere: authors that link to other authors, linked – related topics etc.

The *date* and *time* that an entry was registered is another useful piece of information. Analysis of entries based on date and time, will reveal more or less recent weblogs, more or less active weblogs and authors [20], and topics with short or long lifecycle.

### 3.2 The post and weblog graph

This section illustrates the transition from the post to the weblog graph, through a small example. In the post graph (Figure 1), the links between posts are presented with red arrows. The blue lines indicate the author of each post and next to each post is the topic it refers to. The portion comprises of 11 posts with only 3 hyperlinks between them. Obviously the posts graph is very sparse and existing ranking algorithms will not perform well.

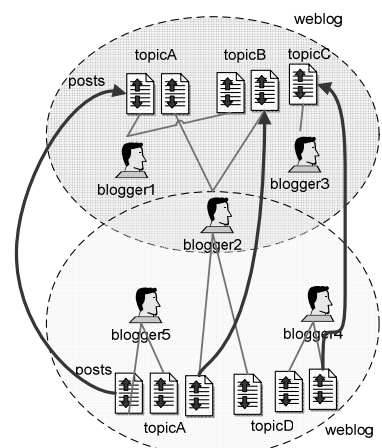

**Figure 1.** Hyperlinks and relations in post level

The weblog graph (Figure 2) aggregates information of the post graph. This produces a denser graph with two nodes connected with multiple edges. The nodes are weblogs comprising of a number of posts, authors and topics. The edges are: aggregated hyperlinks and similarity links based on similarity in authors and topics between the two blogs.

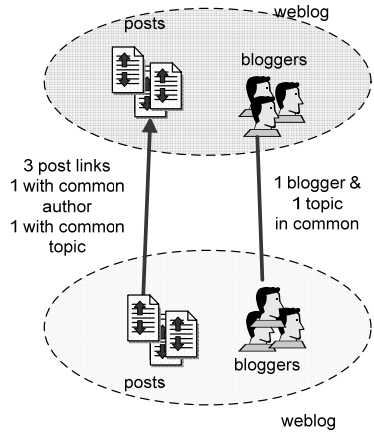

**Figure 2.** Links in weblog level

The proposed algorithm takes into account both link and similarity information in order to estimate the probability of a weblog surfer to follow a link to another weblog. The algorithm prioritizes links to weblogs with similar topics or contributors and assigns greater probability to weblogs which are strongly interconnected.

### 3.3. Database structure

In order to process the dataset we used a relational database. The database schema comprises of three entities, posts, tags and links as depicted in Figure 3. We have two sub-types of outlinks: links to posts and links to news URLs.

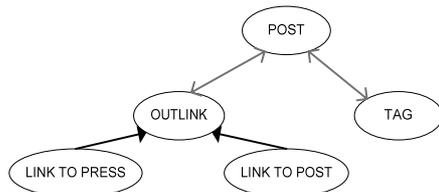

**Figure 3.** The basic entities

The following table and figure summarize the statistics of the dataset we used (using unique URLs).

**Table 1.** Statistics on the graph of posts

| Nodes | | | Edges | | |
|---|---|---|---|---|---|
| Crawled posts URLs | All post URLs | # news' URLs | Links to posts with content | Links to posts | links to news |
| 7,637,399 | **7,967,950** | 212,504 | 331,068 | **2,138,381** | 498,834 |

The first three columns of Table 1 contain information on the nodes of our graph (unique permalinks with content, all permalinks, news URLs). The other three columns provide information on the edges of the graph (links to permalinks with content, links to all permalinks, links to news URLs). By dividing the number of edges to the number of nodes we get approximately 0.27 links per post in the sparse post graph.

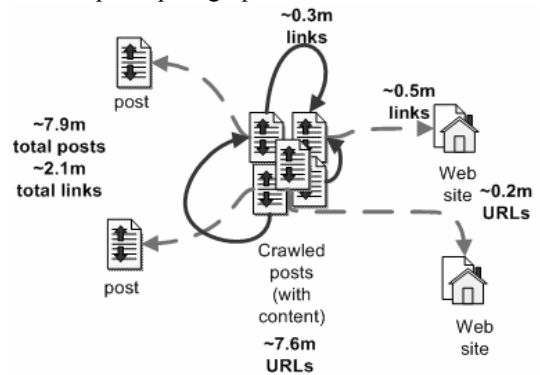

**Figure 4.** The posts graph

Figure 4 presents a sketch of the graph (numbers in millions) in the post level, which gives useful statistics on the number of outlinks from the core set of posts to: web pages, to posts inside the core and post outside of it. The statistics of the weblog graph are presented in table 2.

**Table 2.** Statistics of the weblog graph

| Nodes | Edges | |
|---|---|---|
| unique weblogs | Aggregated links between weblogs | Aggregated links from weblogs to news URLs |
| 1,545,205 | 761,427 | 389,626 |

The second and third column, contain the number of unique links among weblogs and between weblogs and news' URLs. As expected the new graph has significantly fewer nodes (1.5 instead of 7.9 millions) whilst the number of edges decreases in a smaller rate (0.8 instead of 2.1 million). A division between columns 1 and 2 gives a denser graph with an average of 0.49 edges per node.

In order to further increase the number of edges in the graph we generate more implicit links based on similarity in authors and tags. We add a link between two weblogs when they have many authors in common or when their posts have many tags in common. After experimentation we decide on the minimum values for the number of authors and tags. It is on our knowledge that this inclusion strategy is not the best possible solution. However, it is on our next plans to fine tune the graph enhancement process.

For the experiments we created a graph of weblogs. In order to exploit the information conveyed by the links to news URLs we employed the coupling factor [9] used in bibliography research as an indication of similarity between nodes. When two nodes (in our case weblogs) point to many news URLs in common we consider that the nodes are related.

The output of the algorithm, which we explain in the next section, is a ranking of the 1.5 millions weblogs in the dataset. However, the dataset does not contain posts from all these weblogs. So it is very interesting to crawl on the posts of all the weblogs.

## 4. BlogRank

The output of our algorithm is a ranking of all weblogs in the dataset. This overall ranking will be used by our search engine for the presentation of results: matching entries from highly ranked weblogs will be ranked first. In the case of entries from the same weblog, most recent entries are favoured. BlogRank is a generalized approach of Pagerank [19]. The BlogRank of a Weblog A is given by the formula:

$$B(A) = (1-E) + E (FN(U_1 \rightarrow A)* B(U_1) + ... + FN(U_n \rightarrow A)*B(U_n)) \quad (1)$$

where: B(A) is the BlogRank of weblog A,

B(Ui) is the BlogRank of weblog Ui which link to weblog A,

E is a damping factor between 0 and 1 (normally is 0.85)

FN(Un→A) is the possibility of a user that visits weblog n to select weblog A and denotes a factor which shows how much the weblog Un « fancies » weblog A.

The following equation stands:

$$\sum_{j=1}^{t} FN(Uz \rightarrow j) = 1 \quad (2)$$

where:  z is a weblog with t outlinks (to other weblogs)

FN(Uz→j) is the possibility that the user will choose weblog j

If we assume in BlogRank that FN(Uz→j)=1/N where N is the total number of outlinks in weblog z, then we can easily derive the PageRank formula. We strongly believe that a user is not attracted equally by every outlink that exist in post of a given weblog. The most probable case is that the user was driven to a post because she was looking for topic or she is interested for the main subject of the post. It is logical to hypothesize that she is most probably going to continue her quest, by selecting similar post or news. From all the outlinks of weblog z, the significant function FN(Uz→j) favours those posts of the j weblog that:

a) belong to common categories with the weblog z

b) same users have posted as in weblog z

c) link to the same news posts as mentioned in weblog z

Before we apply the BlogRank, we expand the connected graph of the weblogs by adding bidirectional links between the weblogs that share same categories, users and news. Then we apply weights to every

connection.  The utility function that gives the possibility of user to move to weblog j once in z is :

$$FN(U_{z \rightarrow j}) = \frac{F_{z \rightarrow j}}{\sum (F_{z \rightarrow x})} \quad (3a)$$

Where

$$F_{z \rightarrow k} = L_{z \rightarrow k} + w_T * T_{z \rightarrow k} + w_U * U_{z \rightarrow k} + w_N \quad (3b)$$

and

L is the number of links from weblog z to weblog j

T is the number of common tags/categories between z and j

U is the number of users that have posted in both z and j

N is the number of couplings of z and j to news URLs.

$W_T$, $W_U$, $W_N$ are the weights we use in each one of the factors T,U,N respectively.

In our experiments we adjusted BlogRank with the following weights: $W_T$=2, $W_U$=1, $W_N$=3. This is subject to change after experimenting with multiple combinations of the parameters in the utility function. Our aim is to maximize user satisfaction from the search engine results. Although the selected weights are not fine-tuned, the first results we have available (see section 5.3) show that BlogRank outperforms the rest of the algorithms we tested.

In the following we give a short example on how we calculate Function 3b. The probability for a user that reads a post in the bottom weblog to follow a link to a post in the top weblog is depended to the relevance between the two blogs. Function 3b takes into account the number of links between posts of the two blogs ($L_{bottom \rightarrow top}$=3), the number of common tags between posts of the two blogs ($T_{bottom \rightarrow top}$=1), the number of common authors ($U_{bottom \rightarrow top}$=1) and the number of common links to news pages (i.e. $U_{bottom \rightarrow top}$=1).

As a result, the nominator of FN($U_{bottom \rightarrow top}$) will be 9 (3+2*1+1*1+3*1=9). For the denominator we should consider the whole graph and take into account all the links between the bottom weblog and other weblogs.

## 5. Experiments

### 5.1. Extending the graph

By adding implicit links on the graph we increase its density and consequently the performance of the ranking algorithm. Of course, the notion of the web surfer, in which PageRank is based, is more obfuscated if we consider implicit links. In the extended graph, we consider the probability that the reader of a weblog A will move to a related weblog B, even when A and B are not hyperlinked. This transition requires additional steps (i.e. a recommendation engine that suggests "similar" posts), but is feasible with existing technologies [4].

We create implicit links by setting thresholds in the similarity factors. The thresholds have been set up experimentally and possibly need further tuning. Moreover, we adjust the four weights in formula 3 and test three combinations of weights.

The thresholds are as follows:

- The minimum required number of common tags is 3. Very rare tags (those that appeared in less than 500 weblogs) were excluded from the experiment. This results in 1,199,102 tag similarity links between weblogs.

- The minimum required number of common authors is 2. This results in 30,064 author similarity links. Authors with common username (such us "Admin", "John" or "Webmaster") were excluded.

- The adequate coupling factor is set to 2. This results in 2,232,488 news similarity links.

Since many of the implicit links between weblogs overlap the final extended graph contains a total of 4,205,831 distinct edges. We decide to use three different sets of weights ($W_T$, $W_U$, $W_N$) and for L. We distinguish the following three cases:

We set the three weights to 0 and L=1. This means that we collapse the multiple links between two weblogs into a single link and ignore all implicit links. This gives the simple PageRank formula. We refer to this formula as Rank1 (PageRank).

We set the weights to 0 and make L equal to the number of distinct links between posts. We still ignore the implicit links but consider the multiplicity of links. This ranking is an extension of the PageRank algorithm and we call it Rank2 (or XRank). This ranking considers only one type of edges (hyperlinks) in the graph of weblogs but assigns weights based on the number of links between weblogs and is an average solution between PageRank and BlogRank.

Finally after a few experiments we decide on setting the weights to (2, 1, 3) for (common tags, common authors and common links to press articles) respectively, in an attempt to subjectively judge the importance of each type of implicit link. L still represents the cardinality of links. We call the formula Rank3 (BlogRank).

The average number of inlinks per node in the graph employed by PageRank and XRank are 3.61, whilst the average for outlinks is 6.65. The averages in the enhanced graph, which is employed by BlogRank are 17.40 for inlinks and 31.60 for outlinks. The following tables present the first 10 results of every type of ranking. We notice that the weblogs that are ranked with Rank1 have similar order with the ones that were ranked with Rank2. But the order of Rank3 seems different than the previous two rankings.

**Table 3.** The 10 best ranked weblogs of Rank 1 (PageRank)

| Weblog | Rank 1 |
|---|---|
| http://www.boingboing.net | 107.82 |
| http://www.engadget.com | 77.75 |
| http://www.livejournal.com/users/grahame | 77.58 |
| http://www.huffingtonpost.com/theblog | 43.39 |
| http://www.livejournal.com/users/interim32 | 39.67 |
| http://radio.weblogs.com/0001011 | 34.59 |
| http://www.gizmodo.com | 34.36 |
| http://grooveadam.blogspot.com | 33.87 |
| http://www.crooksandliars.com | 30.98 |
| http://www.livejournal.com/users/tyrell | 28.97 |

**Table 4.** The 10 best ranked weblogs of Rank 2 (XRank)

| Weblog | Rank 2 |
|---|---|
| http://www.boingboing.net | 114.02 |
| http://www.engadget.com | 106.11 |
| http://www.livejournal.com/users/grahame | 77.44 |
| http://slashdot.org | 57.32 |
| http://www.huffingtonpost.com/theblog | 47.53 |
| http://www.livejournal.com/users/interim32 | 41.69 |
| http://www.gizmodo.com | 40.80 |
| http://www.cinematical.com | 38.49 |
| http://www.crooksandliars.com | 38.29 |
| http://radio.weblogs.com/0001011 | 35.84 |

**Table 5.** The 10 best ranked weblogs of Rank 3 (BlogRank)

| Weblog | Rank 3 |
|---|---|
| http://nocapital.blogspot.com | 108.38 |
| http://www.livejournal.com/users/pseudomanitou | 91.32 |
| http://tbogg.blogspot.com | 80.31 |
| http://www.feministblogs.org | 76.85 |
| http://www.livejournal.com/users/grahame | 76.68 |
| http://www.boingboing.net | 54.85 |
| http://www.livejournal.com/users/mparent7777 | 49.81 |
| http://www.engadget.com | 49.57 |
| http://blog.blogpulse.com | 47.36 |
| http://www.driko.org | 44.79 |

In the results we notice that there are 139 common weblogs in the first 1000 ranked weblogs of each ranking type. The percentage (13.9%) denotes that the algorithms present different results and rank the weblogs in different order. In order to find the ranking algorithm that mostly satisfies users we setup a search engine for searching posts content (section 5.2) and have a group of users to make queries and evaluate the results without being aware of the ranking algorithm used in every case (section 5.3). Our claim is that the BlogRank outperforms the other rankings in terms of users' satisfaction. And the end of this section we present comparative results that prove our claims.

## 5.2. Evaluation

For the evaluation of results we use the Success Index (SI) metric which was presented in Compass Filter [11]. The basic advantage of Success Index is that it does not require the user to vote for her satisfaction. BlogWave records the posts clicked on by the user, and the order in which they are clicked.

We then evaluate the user's response using Success Index, a number between 0 and 1:

$$SI = \frac{1}{n}\sum_{t=1}^{n}\frac{n-t+1}{d_t * n} \qquad (4)$$

where: **n** is the total number of the posts selected by the user

**dt** is the order in the list of the **t**-th post selected by the user

The SI score rewards the clicking of high items early on. The reverse ranks of the items clicked are weight-averaged, with weights decreasing linearly from 1 down to 1/n with each click. For example, suppose n = 2 and the posts ranked 2 and 10 were clicked. If 2 is clicked first, then the SI score is bigger (27.5%); if it is clicked second, the SI is smaller (17.5%). More controversially, SI penalizes many clicks; for example, the clicking order 2-1-3 has higher score than 1-2-3-4. In the absence of rating (when the user visits the post but does not provide a score) we assign zero score to the post. However, in our experiments we excluded the queries for which we have no user feedback.

During our experiment period, we specifically asked many users to enter and use the BlogWave service in order to have as many evaluation data as possible in the short time given. In an attempt to eliminate subjective bias, the experiment was double-blind since neither the individual users nor we know in advance the ranking method that is used in every query (the method was selected randomly). The use of double-blind test, allows the comparison of ranking methods against different query sets, which is the case in our experiment and generally in web search engines. This information is stored in the database and is used only for the evaluation of user satisfaction.

### 5.3. Setup

The hardware we used was only a simple server PC (Pentium 4, 3.2 GHz, 2Gb memory, 117 Gb and 378Gb hard drives, Microsoft Windows 2003 Server). At the same PC we hosted the search engine, and the database with the corpus of Nielsen BuzzMetrics and the user information. For the experiments we extended the existing search engine of SpiderWave with the BlogWave service (figure 5) for searching the blogs in the dataset. The data was split in two SQL databases: one with the content (34.2 GB) and the second (6.6 GB) with all the entities such as post, tag, outlink and the necessary user information. The MSSearch service was used for the indexing of the content, which needed 8.3 GB to populate the indexes. The response time for each query was between 3 and 90 seconds, depending on the complexity of the query (how common a query term was), the resources usage of the server, and the cached information MSSearch service kept.

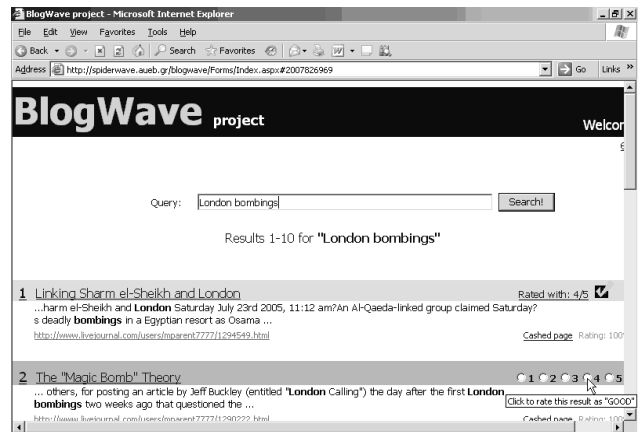

**Figure 5.** Evaluation of the search engine results

The result is a search engine on the content of blogs [17] which presents the results of a search to the user. The results are ranked using one out of three different sets of weights in the ranking formula (Function 3a). For every query the BlogWave executed a database query: at the beginning the 1000 most important posts were asked from the MSSearch service. After that, the posts were ordered by importance of their weblog, depending on the ranking method used for the query. In case of a tie, then the post date is used to order the posts (most recent are presented first). Users evaluated the results by selecting the ranked posts.

The results we present to the user comprise of posts ordered by the ranking of the weblog that they belong. In the case of two posts from the same weblog the most recent post gets a higher rank. Finally we grouped the query results, and computed the average SI score for each one of the three rankings we tested. Below we present the results of our experiment:

**General**
Time period of the Experiment: March 2006 – April 2006 (the experiment is still running, so we continuously collect user evaluation and we are able to tune the ranking formula)
Number of logged-in users: 72
Number of queries asked: 272
Number of queries ranked by the users: 76

**Table 6.** Average SI scores

| Group A) Ranked queries which were based on the Rank 1 - PageRank algorithm |
|---|
| Number of queries:       16 |
| Average Success Index:   0.158 |

| Group B) Ranked queries which were based on the Rank 2 – XRank algorithm |
|---|
| Number of queries:       38 |
| Average Success Index:   0.353 |

| Group C) Ranked queries which were based on the Rank 3-BlogRank algorithm |
| --- |
| Number of queries: 22 |
| Average Success Index: **0.553** |

For group A the coin flipped and determined that the result set of posts will be sorted using the Rank 1 of each weblog that the post was belong to. For group B the engine sorted the presented queries by using the extended PageRank algorithm (Rank2). And finally for group C the results were sorted by using the BlogRank (Rank3). As we can see the average SI scores (0.553) in BlogRank's queries is much higher than the PageRank's (0.158) and XRank's (0.353) equivalent.

Submitting the results of Table 6 to the t-Test statistical analysis method (PRANK1_RANK2 = 0.001 < 0.01 and PRANK2_RANK3 = 0,008 < 0.01) we result that the observed difference between the means is significant, supporting the conclusion that the results of group C are substantial better that the results of the group B and that the results of group B are better than the results of group A. We can safely conclude that the best method is the Rank 3 - BlogRank which appears to considerably improve the quality of the retrieved information.

## 6. Conclusions – Future Work

We have proposed a method for using link graph characteristics and common attributes between the posts to enhance the efficiency of the ranking mechanism for each weblog's importance. Our experimental results are quite encouraging. Much more experimental evaluation of our method, as well as tuning of its parameters is needed.

We developed and tested our method in the context of a very modest fragment of the Weblogging ecosystem. This scaled-down experimentation and prototyping may be an interesting methodology for quickly testing information retrieval ideas, and for expanding the realm of research groups, especially academic groups lacking strong industrial contacts, that are in a position to conduct search engine research. But does our method scale to the entire Blogosphere? First let's check if we could calculate the BlogRank for each weblog for the web. In our case (about 1.5 million weblogs) PageRank took 13 hours to complete, while BlogRank needed 16 hours (23% more time), which is a small difference. So it's safe to assume that the engines of the well known search engines (such as Google, Blogpulse) will use analogically almost the same time to calculate BlogRank, as they would do with Pagerank.

Our effort will focus on the use of objective information for describing the topics of posts and weblogs. For the topic detection process, instead of processing the Content and Tag information of a post, which is controlled by the author, we will use incoming hyperlinks information which is more objective [22].

It is on our plans to process other aspects of the posts graph, more specifically, instead of grouping posts by weblog we plan to group posts "by topic" and "by author", thus forming a graph of interconnected topics and a graph of interconnected authors. The strength of each connection will be based on the number of real links between posts of each topic or author. Both author and topic graphs are directed, strongly connected and have many nodes. Using the biased surfer model we can estimate the probability of a surfer to follow a link to another topic or to another author's post thus revealing the most authoritative authors or topics [18].